\documentclass[12pt,a4paper]{article}

\newcommand{\va}{\varphi}
\newcommand{\mth}{\vartheta}
\newcommand{\pa}{\partial}
\newcommand{\sqp}{\sqrt{p^2 + 1}}

\begin{document}

\begin{flushright}
{ }
\end{flushright}
\vspace{1.8cm}

\begin{center}
 \textbf{\Large BPS Wilson Loop T-dual to \\
Spinning String in $AdS_5 \times S^5$}
\end{center}
\vspace{1.6cm}
\begin{center}
 Shijong Ryang
\end{center}

\begin{center}
\textit{Department of Physics \\ Kyoto Prefectural University of Medicine
\\ Taishogun, Kyoto 603-8334 Japan}
\par
\texttt{ryang@koto.kpu-m.ac.jp}
\end{center}
\vspace{2.8cm}
\begin{abstract}
We use the string sigma model action in $AdS_5 \times S^5$ to 
reconstruct the open string solution ending on the Wilson loop in
$S^3\times R$ parametrized by a geometric angle in $S^3$ and an angle 
in flavor space. Under the interchange of the world sheet space and 
time coordinates and the T-duality transformation with the radial 
inversion, the static open string configuration
associated with the BPS Wilson loop with two equal angle parameters 
becomes a long open spinning string configuration which is produced by 
taking the equal limit of two frequencies for the folded spinning
closed string with two spins in $AdS_5 \times S^5$.
\end{abstract}
\vspace{3cm}
\begin{flushleft}
October, 2013
\end{flushleft}

\newpage
\section{Introduction}

The AdS/CFT correspondence \cite{MM} has more and more revealed the deep
relations between the $\mathcal{N}=4$ super Yang-Mills (SYM) theory
and the string theory in $AdS_5 \times S^5$. A lot of fascinating results
have been accumulated in the computation of the planar observables 
such as the spectrum, Wilson loops and scattering amplitudes \cite{NB}. 

By computing the expectation value of the Wilson loop consisting of a
pair of antiparallel lines from the string theory in $AdS_5 \times S^5$
the effective potential between a pair of heavy W bosons has been
extracted \cite{JMM,RY}. It has been investigated perturbatively in
the gauge theory \cite{ES,ESZ} and by the strong coupling expansion
in the string theory \cite{FGT,DGT,CHR,VF}(see also \cite{MKA}). 

For the circular 1/2 BPS Wilson loop its expectation value evaluated in 
the string theory has been reproduced by performing the 
resummation of ladder diagrams in the $\mathcal{N}=4$ SYM theory 
\cite{ESZ,DG} and using the localization arguments \cite{VP}. 
The lower supersymmetric Wilson loops on a two-sphere $S^2$ embedded into
the $R^4$ spacetime have been analyzed by finding the corresponding
open string solutions as well as by reducing a purely perturbative
calculation in the soluble bosonic 2d Yang-Mills 
on the sphere \cite{DGR,KZ,GRT,VPE}.

In \cite{DF} the effective potential between a generalized quark
antiquark pair has been computed by studying a family of Wilson loops in
$S^3 \times R$ which are parametrized by two angle parameters $\phi$ and
$\theta$. The quark and antiquark lines are extended along the time 
direction and are separated by an angle $\pi - \phi$ on $S^3$.
The parameter $\theta$ is the relative orientation of the extra coupling
to the scalar field for the quark and the antiquark.
Through the plane to cylinder transformation, the two lines in 
$S^3 \times R$ map into two half-lines with a cusp of angle $\pi - \phi$
in $R^4$, and the potential energy of static quark and antiquark is 
identical with the cusp anomalous dimension.
The Wilson loop in $S^3 \times R$ with the Minkowski signature 
interpolates smoothly between the 1/2 BPS two antipodal lines at 
$\phi = 0$ and the coincident two antiparallel lines at $\phi = \pi$,
while the two Wilson lines with a cusp in $R^4$ between the 1/2 BPS 
one straight line and the coincident two lines with a cusp of zero angle.

In the weak coupling expansion for the $\mathcal{N}=4$ SYM theory
the effective potential  has been computed at one-loop order \cite{DGO}
and at two-loop order for $\phi = 0$ \cite{MOS}, for $\phi \ne 0$
\cite{DF}. In the semiclassical expansion for the string theory by using 
the Nambu-Goto string action, the effective potential has been evaluated
at leading order \cite{DGO,DGR} and at one-loop order \cite{DF}.

An exact formula for the Bremsstrahlung function of the cusp anomalous
dimension for small values of $\phi$ and $\theta$ has been found by
relating the cusp anomalous dimension to the localization result of
certain 1/8 BPS circular Wilson loops \cite{CH} (see also \cite{FGL}).
The first two terms in the weak coupling and the strong coupling 
expansions of the exact formula agree with the results of the 
corresponding effective potential in \cite{DF}.
The three loop term in the weak coupling expansion has been produced by 
the explicit three loop computation \cite{CHM} and the three 
loop expansion of the TBA equations \cite{CMS,ND}.
There have been further studies about the cusp anomalous dimension
associated with the cusped Wilson lines \cite{GS,BZ,FBT}.

Suggested by a striking similarity between the Bremsstrahlung function
\cite{CH} of the cusp anomalous dimension in the small angle limit
and the slope function \cite{BB} found in the small spin limit of the 
$AdS_5$ folded string energy, there has been a construction of a possible
relation between small ( nearly point-like ) closed strings in $AdS_5$
and long open strings ending at the boundary which correspond to nearly
straight Wilson lines \cite{KT}. Through the T-duality along the boundary
directions of Lorentzian $AdS_5$ in the Poincare coordinates 
together with the radial inversion $z \rightarrow 1/z$ 
\cite{RKT,LAM,RTW} and the interchange of space and 
time coordinates of the Minkowski world sheet,  
the world sheet of small closed string is related with
the open string surface ending on wavy line representing small-velocity
``quark" trajectory at the boundary. This open string solution 
corresponds to the small-wave open string solution in \cite{AM} which
ends on a time-like near BPS Wilson loop differing by small fluctuations
from a straight line. Further from the computation of the one-loop 
fluctuations about the classical small-wave open string solution, 
the one-loop correction to the energy radiated by the end-point of
a string has been evaluated \cite{BT} to be consistent with the subleading
term in the strong coupling expansion of the Bremsstrahlung function
in \cite{CH}.

Instead of the Nambu-Goto action we will use the string sigma model 
action in the global coordinates to reconstruct the open string solution
ending at the boundary which is associated with the two antiparallel 
Wilson lines in $S^3 \times R$ parametrized by two angle parameters
$\phi$ and $\theta$ \cite{DF}. We will express the Minkowski open string
solution associated with the BPS Wilson loop \cite{KZ} specified by 
$\phi = \pm\theta$ in terms of the Poincare coordinates.
We will first make the flip of world sheet coordinates and secondly
perform the T-duality along the boundary 
directions with the radial inversion
$z \rightarrow 1/z$ to see what kind of string configuration appears
and examine how the BPS condition $\phi = \pm\theta$ is encoded in the
transformed string configuration.

\section{The open string solution for the BPS Wilson loop}

We consider the classical open string solution in $AdS_5 \times S^5$
ending on the Wilson loop at the boundary which
interpolates smoothly between the 1/2 BPS two antipodal lines
and the coincident two antiparallel lines \cite{DF}.
The Wilson loop for the $\mathcal{N}=4$ SYM theory in $S^3 \times R$ is
given by
\begin{equation}
W = \frac{1}{N}\mathrm{Tr}\mathcal{P}\mathrm{exp}\left[ \oint 
(iA_{\mu} {\dot{x}}^{\mu} + \Phi_I \Theta^I |\dot{x}|) ds \right]
\end{equation}
and characterized by two parameters $\phi$ and $\theta$, where the
loop is made of two lines separated by an angle $\pi - \phi$ along the big
circle on $S^3$, and $\theta$ specifies the coupling to the scalars 
$\Phi_I$. By describing the angle along the big circle by $\va$ and 
the time by $t$ we parametrize the two lines extending 
to the future and the past time directions as
\begin{eqnarray}
t &=& s, \;\; \va = \frac{\phi}{2}, \;\; \Theta^1 = \cos 
\frac{\theta}{2}, \;\;  \Theta^2 = \sin \frac{\theta}{2}, 
\nonumber \\
t &=& -s', \;\; \va = \pi - \frac{\phi}{2}, \;\; \Theta^1 = \cos 
\frac{\theta}{2}, \;\;  \Theta^2 = -\sin \frac{\theta}{2}.
\label{ts}\end{eqnarray}
The general open string solution with two arbitrary angles
$\phi$ and $\theta$ was constructed by using the Nambu-Goto action 
\cite{DF}. 

We rederive the open string solution associated with the BPS Wilson loop
with $\phi = \pm\theta$ by using the conformal gauge for the string sigma
model in the global coordinates.
For a static open string in $AdS_3 \times S^1$ with metric
\begin{equation}
ds^2 = - \cosh^2 \rho dt^2 + d\rho^2 + \sinh^2\rho d\va^2
+ d\mth^2
\end{equation}
we make the following ansatz
\begin{equation}
t = \tau, \;\; \rho = \rho(\sigma), \;\; \va = \va(\sigma), \;\;
\mth = \mth(\sigma)
\end{equation}
with the Minkowski signature in the world sheet.
The Virasoro constraint yields
\begin{equation}
-\cosh^2 \rho + (\pa_{\sigma}\rho)^2 + \sinh^2\rho (\pa_{\sigma}\va)^2
+ (\pa_{\sigma}\mth)^2 = 0,
\label{vi}\end{equation}
which reads  
\begin{equation}
\cosh^2 \rho = (\pa_{\sigma}\va)^2\left[ (\pa_{\va}\rho)^2 + \sinh^2\rho
+ (\pa_{\va}\mth)^2 \right].
\end{equation}
We make a choice of sign as
\begin{equation}
\pa_{\sigma}\va = \frac{\cosh\rho}{\sqrt{ (\pa_{\va}\rho)^2 + \sinh^2\rho
+ (\pa_{\va}\mth)^2 }}.
\label{vas}\end{equation}

The equation of motion for $\va$
\begin{equation}
\pa_{\sigma}( \sinh^2\rho \pa_{\sigma}\va ) = 0
\label{vsi}\end{equation}
gives one integral of motion $p$ as
\begin{equation}
\sinh^2\rho \pa_{\sigma}\va  =  \frac{1}{p},
\label{psh}\end{equation}
which is expressed through (\ref{vas}) as
\begin{equation}
\frac{\sinh^2 \rho \cosh\rho}{\sqrt{ (\pa_{\va}\rho)^2 + \sinh^2\rho
+ (\pa_{\va}\mth)^2 }} =  \frac{1}{p},
\label{prh}\end{equation}
where we choose $p$ as a positive parameter.
The equation of motion for $\mth$ 
\begin{equation}
\pa_{\sigma}^2 \mth = 0
\label{ths}\end{equation}
gives a solution $\mth = J\sigma$ and the other integal of motion $J$
defined by $\pa_{\sigma} \mth = J$ is  described through
(\ref{vas}) as
\begin{equation}
\frac{\pa_{\va}\mth \cosh\rho}{\sqrt{ (\pa_{\va}\rho)^2 + \sinh^2\rho
+ (\pa_{\va}\mth)^2 }} = J.
\label{vj}\end{equation}
We combine (\ref{vj}) with (\ref{prh}) to have
\begin{equation}
\frac{\pa_{\va}\mth}{\sinh^2\rho} = pJ.
\label{pj}\end{equation}

The remaining equation of motion for $\rho$ is 
\begin{equation}
\pa_{\sigma}^2 \rho - \cosh\rho\sinh\rho - 
\cosh\rho\sinh\rho( \pa_{\sigma}\va )^2 = 0,
\label{rem}\end{equation}
whose first integral is contained in (\ref{vi}). Indeed differentiating
(\ref{vi}) with respect to $\sigma$ we have
\begin{equation}
\pa_{\sigma}\rho ( \pa_{\sigma}^2 \rho - \cosh\rho\sinh\rho ) + 
\pa_{\sigma}\va  ( \sinh^2\rho\pa_{\sigma}^2 \va + \cosh\rho\sinh\rho
\pa_{\sigma}\rho\pa_{\sigma}\va ) + \pa_{\sigma}\mth \pa_{\sigma}^2\mth
= 0,
\end{equation}
which yields (\ref{rem}) through (\ref{vsi}) and (\ref{ths}).
The expressions such as (\ref{prh}), (\ref{vj}) and (\ref{pj}) were 
presented in ref. \cite{DF} for the general $\phi \ne \theta$ case with
two arbitrary parameters $J$ and $p$ where $p$ is defined to have
opposite sign from ours.

Here we restrict ourselves to the $J = \pm1/p$ case, that is , the 
$\phi = \pm\theta$ case. Substitution of (\ref{pj}) into (\ref{vj}) 
generates a differential equation for $\rho$ 
\begin{equation}
(\pa_{\va}\rho)^2 = p^2 \cosh^2\rho\sinh^4\rho - \sinh^4\rho
- \sinh^2\rho,
\label{rhv}\end{equation}
which becomes through (\ref{psh}) to be
\begin{equation}
(\pa_{\sigma}\rho)^2 = \frac{\cosh^2\rho( p^2 \cosh^2\rho - (p^2 + 1) )}
{p^2\sinh^2\rho}.
\label{sir}\end{equation}
The turnning point $\rho_0 \; (\rho_0 \le \rho)$ of open string
is specified by
\begin{equation}
\cosh^2 \rho_0 = \frac{p^2 + 1}{p^2}.
\end{equation}
Thus we have dealt with the string sigma model action to recover
the same relevant equations as constructed by using Nambu-Goto string
action in ref. \cite{DF}. Using the string sigma model for the open string
in $AdS_3 \times S^3$ with a spin in $S^3$ the first integrals for
equations of motion have been presented for the $\theta \ne 0,
\phi = 0$ case \cite{CMS} and for the $\phi \ne \theta$ case \cite{GS}.
Without solving the relevant equations to obtain the string profile
the angles and charges have been directly extracted from them.

 We are interested in the explicit expression of
solution for (\ref{sir}) specified by a single parameter $p$, which can 
be read off by putting $J = \pm1/p$ in the general solution for $\rho$
with the two parameters $J$ and $p$ in ref. \cite{DF}.
It is convenient to rescale the world sheet coordinates 
$\tau$ and $\sigma$ in the same way
\begin{equation}
\tau \rightarrow \frac{p}{\sqp}\tau, \hspace{1cm}
\sigma \rightarrow \frac{p}{\sqp}\sigma.
\end{equation}
By integrating the rescaled equation of (\ref{sir}) 
under a condition $\rho = \rho_0$ at $\sigma = 0$ 
we have the following open string solution 
\begin{equation}
t = \frac{p}{\sqp}\tau, \;\; \mth = \pm\frac{1}{\sqp}\sigma, \;\;
\cosh \rho = \frac{\sqp}{p} \frac{1}{\cos \sigma}, \nonumber \\
\label{sso}\end{equation}
where the ranges of the world sheet coordinates are
\begin{equation}
- \frac{\pi}{2} \le \sigma \le \frac{\pi}{2}, \hspace{1cm}
- \infty < \tau < \infty.
\end{equation}
At the open string ends $\sigma = \pm \pi/2$, the radial coordinate $\rho$
diverges such that the string touches the boundary of the $AdS_3$ space.
The angle $\mth$ changes in $-\theta/2 \le \mth \le \theta/2$,
\begin{equation}
\theta = \pm\frac{\pi}{\sqp}.
\end{equation}

Substitution of the third equation in (\ref{sso}) into the rescaled 
equation of (\ref{psh}) yields
\begin{equation}
\pa_{\sigma}\va = \frac{1}{\sqp} \frac{p^2\cos^2\sigma}{p^2 + 1
-p^2\cos^2\sigma}.
\end{equation}
The integration under a condition 
\begin{equation}
\va(\sigma=-\frac{\pi}{2}) = \frac{\phi}{2} = \pm\frac{\theta}{2}
= \frac{\pi}{2\sqp}
\end{equation}
as (\ref{ts}) gives a solution for $-\pi/2 \le \sigma \le 0$
\begin{equation}
\va = - \frac{1}{\sqp}\left[ \sigma - \sqp \tan^{-1}\left( 
\frac{1}{\sqp} \tan \left( \sigma + \frac{\pi}{2} \right) \right) \right].
\label{fva}\end{equation}
For $0 \le \sigma \le \pi/2$ we need to analytically continue the 
solution. In that region it is given by
\begin{equation}
\va = \pi -\frac{\sigma}{\sqp} - \tan^{-1}\left( \frac{1}{\sqp}
\tan\left( \frac{\pi}{2} - \sigma \right) \right),
\label{vap}\end{equation}
which becomes $\pi - \phi/2$ at $\sigma = \pi/2$ as (\ref{ts}).
 At $p = 0$, the angle $\phi$ becomes $\pi$ such 
that the two antiparallel lines are conincident, while at the infinite $p$
it vanishes such that the two lines are antipodal.

\section{T-duality to the BPS Wilson loop}

We express the string configuration in terms of the Poincare coordinates.
The embedding coordinates $X_M \; (M = 0, \cdots,5)$
for the Lorentzian $AdS_5$ space are described by the global coordinates
$(t, \rho, \psi, \va_1, \va_2)$
\begin{eqnarray}
X_{-1} + iX_0 &=& \cosh \rho e^{it}, \;\; X_1 + iX_2 = \sinh \rho 
\cos\psi e^{i\va_1}, \;\;  X_3 + iX_4 = \sinh \rho \sin\psi e^{i\va_2},
\nonumber \\
X_MX^M &=& -X_{-1}^2 + X_{\mu}X^{\mu} + X_4^2 = -1.
\label{glo}\end{eqnarray}
These coordinates are related with the Poincare coordinates 
$(z, x_{\mu}), \; ds^2 = z^{-2}(dz^2 + dx_{\mu}dx^{\mu})$,
\begin{equation}
X_{\mu} = \frac{x_{\mu}}{z}, \;\; 
X_4 = \frac{-1 + z^2 + x_{\mu}x^{\mu}}{2z}, \;\;
X_{-1} = \frac{ 1 + z^2 + x_{\mu}x^{\mu}}{2z},
\label{poi}\end{equation}
which also define the light-like coordinates
\begin{equation}
X_- = X_{-1} - X_4 = \frac{1}{z}, \;\; 
X_+ = X_{-1} + X_4 = \frac{z^2 + x_{\mu}x^{\mu}}{z}.
\end{equation}

In the $AdS_3$ space specified by $\va_1 = \va, \; \psi = 0$
or $X_3 = X_4 = 0$ we use the static string solution (\ref{sso})
 together with (\ref{glo}) and (\ref{poi}) to derive
\begin{equation}
z = \frac{p}{\sqp} \frac{\cos\sigma}{\cos \frac{p}{\sqp}\tau}, \;\;
x_0 = \tan \frac{p}{\sqp}\tau
\label{zxt}\end{equation}
and
\begin{equation}
(x_1, \; x_2) = \frac{\sqrt{p^2 \sin^2\sigma + 1}} {\sqp    
\cos \frac{p}{\sqp}\tau }(\cos\va, \; \sin\va).
\label{xxv}\end{equation}
Indeed an equation $z^2 + x_{\mu}x^{\mu} = 1$, that is, $X_4 = 0$
is satisfied. 

In the region $-\pi/2 \le \sigma \le 0$ we substitute (\ref{fva})
into (\ref{xxv}) to obtain 
\begin{eqnarray}
x_1 &=& \frac{1}{\cos \frac{p}{\sqp}\tau} \left( - \sin\sigma 
\cos\frac{\sigma}{\sqp} + \frac{1}{\sqp}\cos\sigma \sin\frac{\sigma}{\sqp}
\right), \nonumber \\
x_2 &=& \frac{1}{\cos \frac{p}{\sqp}\tau} \left(\frac{1}{\sqp}\cos\sigma 
\cos\frac{\sigma}{\sqp} +  \sin\sigma 
\sin\frac{\sigma}{\sqp} \right). 
\label{xx}\end{eqnarray}
They are combined to be 
\begin{equation}
x_1 + ix_2 = \frac{1}{\cos \frac{p}{\sqp}\tau}\left( - \sin\sigma 
+ i \frac{\cos\sigma}{\sqp} \right) e^{-i\sigma/\sqp }.
\label{com}\end{equation}
In the region $0 \le \sigma \le \pi/2$ using (\ref{vap})
we compute $x_1$ and $x_2$ to derive the same 
expression as (\ref{xx}) so that we regard (\ref{xx}) as  a solution
in the whole interval $-\pi/2 \le \sigma \le \pi/2$. 
Thus the string configuration is described by $z$ and $x_0$ in (\ref{zxt})
together with the single expression (\ref{xx}).

In the Poincare coordinates the equations of motion for $x_{\mu}$ read
\begin{equation}
\pa_{\tau}\left(\frac{\pa_{\tau}x_{\mu}}{z^2} \right) - 
\pa_{\sigma}\left(\frac{\pa_{\sigma}x_{\mu}}{z^2} \right) = 0,
\;\; \mu = 0, 1, 2.
\label{xz}\end{equation}
Plugging the expressions of $z, x_0$ in (\ref{zxt}) and $x_1, x_2$ in
(\ref{xx}) into (\ref{xz}) we can confirm that they are satisfied.
The equation of motion for $z$ is given by 
\begin{equation}
\pa_{\tau}\left(\frac{\pa_{\tau}z}{z^2} \right) - 
\pa_{\sigma}\left(\frac{\pa_{\sigma}z}{z^2} \right) = \frac{1}{z^3}
\left( - \pa_a x_0\pa^a x_0 + \pa_a x_i\pa^a x_i + \pa_a z\pa^a z \right)
\label{zx}\end{equation}
with $i =1, 2$ and $a =\tau, \sigma$. We confirm that (\ref{zx}) is indeed
satisfied by deriving
\begin{eqnarray} 
\pa_a x_i\pa^a x_i &=& \frac{p^2}{p^2 + 1} \frac{1}{\cos^4 
\frac{p}{\sqp}\tau} \biggl[ - \left( \sin^2\sigma 
+ \frac{\cos^2\sigma}{p^2 + 1} \right)\sin^2 \frac{p}{\sqp}\tau
 \nonumber \\
&+& \frac{p^2}{p^2 + 1}\cos^2\sigma \cos^2 \frac{p}{\sqp}\tau \biggr]
\label{xp}\end{eqnarray}
and showing the RHS of (\ref{zx}) to be 
\begin{equation}
\frac{p^2}{z^3( p^2 + 1)}\frac{1}{\cos^4 \frac{p}{\sqp}\tau}
\left( 1 + \frac{p^2}{p^2 + 1}\cos^2\sigma + \sin^2\sigma \right)
\cos^2 \frac{p}{\sqp}\tau,
\label{zp}\end{equation}
where the $\sin^2 p\tau/\sqp$ term vanished.

In \cite{HOS} a family of closed 
string solutions on $R \times S^3$ subspace
of $AdS_5 \times S^5$ were constructed by using the $\tau \leftrightarrow
\sigma$ flip which maps spinning closed string states with large spins
to oscillating states with large winding numbers.
There was a prescription \cite{KT} that starting from the closed string 
solutions in the bulk of AdS space we perform the T-duality along the
boundary directions in the Poincare coordinates of AdS space 
with the radial inversion and relax
the condition of periodicity in $\sigma$ to interchange
$\tau$ and $\sigma$ for constructing the open string solutions ending at 
the boundary which are associated with the Wilson loops. 

Here we proceed in the inverse direction. Starting from the open string
solution (\ref{zxt}) with (\ref{com}) and $\mth = \pm\sigma/\sqp$ 
which is associated with the BPS Wilson loop we
relax the ranges of the world sheet coordinates and interchange
$\tau$ and $\sigma$ to have
\begin{eqnarray}
\tilde{z} &=& \frac{p}{\sqp} \frac{\cos\tau}{\cos \frac{p}{\sqp}\sigma},
 \hspace{1cm} \tilde{x}_0 = \tan \frac{p}{\sqp}\sigma, \nonumber \\
\tilde{x}_1 + i\tilde{x}_2 &=& \frac{1}{\cos \frac{p}{\sqp}\sigma}\left(
 - \sin\tau + i \frac{\cos\tau}{\sqp} \right) e^{-i\tau/\sqp },
\nonumber \\
\tilde{\mth} &=& \pm\frac{\tau}{\sqp}.
\label{til}\end{eqnarray}

If we make the T-duality along  the boundary directions and 
the inversion of the radial coordinate for some unknown string
configuration $z, x_{\mu}$ as
\begin{equation}
\pa_{\tau}\tilde{x}_{\mu} = - \frac{1}{z^2}\pa_{\sigma}x_{\mu},
\;\; \pa_{\sigma}\tilde{x}_{\mu} = - \frac{1}{z^2}\pa_{\tau}x_{\mu},
\;\; \tilde{z} = \frac{1}{z},
\label{tdu}\end{equation}
then we suppose to obtain the previous string configuration
$\tilde{z}, \tilde{x}_{\mu}$ in (\ref{til}). 
Rewriting (\ref{tdu}) as
\begin{eqnarray}
z &=& \frac{1}{\tilde{z}} = \frac{\sqp}{p} 
\frac{\cos \frac{p}{\sqp}\sigma}{\cos\tau}, 
\label{zz} \\
\pa_{\sigma}x_{\mu} &=& - \frac{1}{\tilde{z}^2}
\pa_{\tau}\tilde{x}_{\mu}, \hspace{1cm}
\pa_{\tau}x_{\mu} = - \frac{1}{\tilde{z}^2}\pa_{\sigma}\tilde{x}_{\mu}
\label{dut}\end{eqnarray}
we substitute the $\tau \leftrightarrow \sigma$ flip solution (\ref{til})
into (\ref{dut}) to have the differential equations for the combination
$x_1 + ix_2$ 
\begin{eqnarray}
\pa_{\sigma}( x_1 + ix_2 ) &=&  
\frac{\cos \frac{p}{\sqp}\sigma}{\cos\tau}e^{-i\tau/\sqp },
\nonumber \\
\pa_{\tau}( x_1 + ix_2 ) &=&  -\frac{\sqp}{p}
\frac{\sin \frac{p}{\sqp}\sigma}{\cos^2\tau}
\left( - \sin\tau + i \frac{\cos\tau}{\sqp} \right)e^{-i\tau/\sqp },
\end{eqnarray}
which can be solved by
\begin{equation}
x_1 + ix_2 =  \frac{\sqp}{p}\frac{\sin \frac{p}{\sqp}\sigma}{\cos\tau}
e^{-i\tau/\sqp }.
\label{xxt}\end{equation}
The other $\mu = 0$ component is obtained by
\begin{equation}
x_0 = - \frac{\sqp}{p}\tan \tau.
\label{xo}\end{equation}

It can be also confirmed that the obtained 
string configuration expressed by
(\ref{xxt}), (\ref{xo}) and (\ref{zz}) indeed satisfies the string 
equations of motion (\ref{xz}) and  (\ref{zx}) in the same way as
shown previously for the starting static open string 
solution. Here for convenience
we write down two relevant equations corresponding to
(\ref{xp}) and (\ref{zp}) respectively
\begin{eqnarray}
\pa_a x_i\pa^a x_i &=& \frac{p^2 + 1}{p^2}\frac{1}{\cos^4 
\tau} \biggl[ -  \sin^2\frac{p}{\sqp}\sigma \left(  \sin^2\tau +
\frac{\cos^2\tau}{p^2 + 1} \right) \nonumber \\
&+& \frac{p^2}{p^2 + 1}\cos^2
\frac{p}{\sqp}\sigma \cos^2\tau \biggr], 
\end{eqnarray}
and 
\begin{equation}
\frac{p^2 + 1}{z^3p^2}\frac{1}{\cos^4 \tau}\left( 1 + \frac{p^2}{p^2 + 1}
\cos^2\frac{p}{\sqp}\sigma + \frac{p^2 - 1}{p^2 + 1}
\sin^2\frac{p}{\sqp}\sigma \right)\cos^2\tau.
\end{equation}

Under the sign change of $\tau$ the string solution in $AdS_3 \times S^1$
becomes
\begin{eqnarray}
z &=& \frac{\sqp}{p}\frac{\cos \frac{p}{\sqp}\sigma}{\cos\tau},
\hspace{1cm}  x_0 =  \frac{\sqp}{p}\tan \tau,
\nonumber \\
x_1 + ix_2 &=&  \frac{\sqp}{p}\frac{\sin \frac{p}{\sqp}\sigma}{\cos\tau}
e^{i\tau/\sqp }, \hspace{1cm} \mth = \mp \frac{\tau}{\sqp}.
\end{eqnarray}
which, however obeys
\begin{equation}
z^2 - x_0^2 + x_1^2 + x_2^2 = \frac{ p^2 + 1}{p^2}.
\end{equation}
In terms of the embedding coordinates the string solution is
expressed as
\begin{eqnarray}
X_0 &=& \frac{\sin \tau}{\cos\frac{p}{\sqp}\sigma}, \;\; 
X_1 + i X_2 = \tan\frac{p}{\sqp}\sigma e^{i\tau/\sqp }, 
\nonumber \\
X_+ &=&  \frac{\sqp}{p} \frac{\cos\tau}{\cos\frac{p}{\sqp}\sigma}, \;\;
X_- =  \frac{p}{\sqp} \frac{\cos\tau}{\cos\frac{p}{\sqp}\sigma}.
\label{emx}\end{eqnarray}
This T-dual string solution has nonvanishing $X_4$.

Now we make a particular SO(2,4) transformation in the
$(X_{-1}, X_4)$ plane, that is, a dilatation transformation
for (\ref{emx}) as
\begin{equation}
X_+ \rightarrow \lambda X_+ , \;\; X_- \rightarrow \frac{1}{\lambda} X_-,
\;\; X_{\mu}: \; \mathrm{invariant}
\end{equation}
with $\lambda = p/\sqp$ to have
\begin{equation}
X_+ = X_- = \frac{\cos\tau}{\cos\frac{p}{\sqp}\sigma},
\end{equation}
which implies $X_4 = 0$. Accordingly the string solution  
in the Poincare coordinates is rescaled as
$z \rightarrow \lambda z, \; x_{\mu} \rightarrow \lambda x_{\mu}$
\begin{equation}
z = \frac{\cos \frac{p}{\sqp}\sigma}{\cos\tau}, \;\;
 x_0 = \tan \tau, \;\;
x_1 + ix_2 = \frac{\sin \frac{p}{\sqp}\sigma}{\cos\tau}e^{i\tau/\sqp },
\end{equation}
which obeys $z^2 + x_{\mu}^2 = 1$ to stay in $AdS_3$.

Further the T-dual string solution in $AdS_3 \times S^1$  
is expressed in terms of the global coordinates as
\begin{eqnarray}
t &=& \tau, \;\; \cosh \rho = \frac{1}{\cos \frac{p}{\sqp}\sigma},
\;\; \va = \frac{1}{\sqp}\tau, \nonumber \\
\mth &=& \mp \frac{1}{\sqp}\tau.
\label{sgl}\end{eqnarray}
The string reaches the $AdS_3$ boundary at
\begin{equation}
\sigma = \pm \frac{\sqp}{p}\frac{\pi}{2}.
\end{equation}
We see that the dilatation transformation plays an important role
to derive a compact and consistent real string solution in the 
global coordinates.

Here we consider the folded spinning closed string 
in $AdS_3 \times S^1$ \cite{FT}
\begin{eqnarray}
t &=& \tau, \;\; \va = \omega\tau, \;\; \mth = \nu \tau,
\label{spi} \\
\rho &=& \rho(\sigma) = \rho(\sigma + 2\pi),
\end{eqnarray}
where $\omega$ and $\nu$ are two frequencies associated with
two spins in $AdS_3$ and in $S^1$ respectively.
The equation of motion for $\rho$ is
\begin{equation}
\pa_{\sigma}^2\rho = ( 1 - \omega^2 ) \sinh\rho\cosh\rho.
\label{erq}\end{equation}
The Virasoro constraint leads to
\begin{equation}
(\pa_{\sigma}\rho)^2 = ( 1 - \nu^2 )\cosh^2\rho - (\omega^2 - \nu^2 )
\sinh^2\rho.
\label{rvi}\end{equation}

In the special $\omega = \mp \nu$ case
the solution for (\ref{rvi}) is given by
\begin{equation}
\cosh\rho =  \frac{1}{\cos \sqrt{1 - \nu^2}\sigma},
\label{coh}\end{equation}
which satisfies the equation of motion for $\rho$, (\ref{erq}).
This string solution is regarded as an effectively long open string
configuration which is stretched along the radial direction of $AdS_3$
to the boundary at $\sigma = \pm \pi/2\sqrt{1 - \nu^2}$.
The specific choice $\nu = \mp1/\sqp$ for (\ref{spi}) and (\ref{coh})
produces the same string solution as the T-dual string configuration
(\ref{sgl}) in $AdS_3 \times S^1$. Thus the T-dual 
string solution can be viewed as a special equal
limit of two frequencies for the folded 
spinning closed string in $AdS_3 \times S^1$. It is noted that
the BPS condition $\phi = \pm\theta$ in the starting static open string
solution becomes the equal condition of two frequencies $\omega = \mp \nu$
in the T-dual spinning string solution.

\section{Conclusion}

Using the string sigma model in the global coordinates on
the Lorentzian $AdS_3 \times S^1$ space we have 
reconstructed the static open string 
configuration ending on the BPS Wilson loop \cite{DF}, which consists 
of the two antiparallel lines specified by two equal
angles $\phi = \pm\theta$. We have expressed the open string solution
in terms of the Poincare coordinates and have used the prescription of
ref. \cite{KT} to make the flip of the Minkowski world sheet coordinates
and perform the T-duality transformation along the boundary directions
with the radial inversion. By solving the T-duality equation we have 
derived a compact expression of solution, which stays out of the 
$AdS_3$ space. We have observed that the  dilatation transformation
plays an important role to put the T-dual string configuration back
into the $AdS_3$ space.

We have demonstrated that when the T-dual string solution 
described by the Poincare coordinates in 
$AdS_3 \times S^1$ is expressed in terms of the global coordinates it 
produces a suggestive expression which turns out to be the spinning 
open string configuration derived by taking the equal limit of
two frequencies for the folded spinning closed string with two 
spins in $AdS_3 \times S^1$. We have observed that the BPS condition
$\phi = \pm\theta$ for the starting BPS Wilson loop corresponds to 
the equal condition of two frequencies associated with two spins
for the T-dual spinning string solution.


\begin{thebibliography}{99}
\bibitem{MM} J.M. Maldacena, ``The large N limit of superconformal
field theories and supergravity," Adv. Theor. Math. Phys. \textbf{2}
(1998) 231 [arXiv:hep-th/9711200]; S.S. Gubser, I.R. Klebanov and 
A.M. Polyakov, ``Gauge theory correlators from non-critical 
string theory," Phys. Lett. \textbf{B428} (1998) 105 
[arXiv:hep-th/9802109]; E. Witten, 
``Anti-de Sitter space and holography,"
Adv. Theor. Math. Phys. \textbf{2} (1998) 253 [arXiv:hep-th/9802150].
\bibitem{NB} N. Beisert $et\, al$., ``Review of AdS/CFT 
integrability: An overview," Lett. Math. Phys. \textbf{99} (2012)
3 [arXiv:1012.3982[hep-th]].
\bibitem{JMM} J.M. Maldacena, ``Wilson loops in large N 
field theories," Phys. Rev. Lett. \textbf{80} (1998) 4859
 [hep-th/9803002].
\bibitem{RY} S.-J. Rey and J.-T. Yee, ``Macroscopic strings as heavy
quarks in large N gauge theory and anti-de Sitter supergravity,"
Eur. Phys. J. \textbf{C22} (2001) 379 [hep-th/9803001].
\bibitem{ES} J.K. Erickson, G.W. Semenoff, R.J. Szabo and K. Zarembo,
``Static potential in $\mathcal{N}=4$ supersymmetric Yang-Mills
theory," Phys. Rev. \textbf{D61} (2000) 105006 [hep-th/9911088].
\bibitem{ESZ} J.K. Erickson, G.W. Semenoff and K. Zarembo,
``Wilson loops in $\mathcal{N}=4$ supersymmetric Yang-Mills
theory," Nucl. Phys. \textbf{B582} (2000) 155 [hep-th/0003055].
\bibitem{FGT} S. Forste, D. Ghoshal and S. Theisen, ``Stringy 
corrections to the Wilson loop in $\mathcal{N}=4$ super Yang-Mills
theory," JHEP \textbf{9908} (1999) 013 [hep-th/9903042].
\bibitem{DGT} N. Drukker, D.J. Gross and A.A. Tseytlin,
``Green-Schwarz string in $AdS_5\times S^5$: Semiclassical
partition function," JHEP \textbf{0004} (2000) 021 [hep-th/0001204].
\bibitem{CHR} S.-x. Chu, D. Hou and H.-c. Ren,``The subleading
term of the strong coupling expansion of the heavy-quark
potential in a $\mathcal{N}=4$ super Yang-Mills vacuum,"
JHEP \textbf{0908} (2009) 004 [arXiv:0905.1874[hep-th]].
\bibitem{VF} V. Forini, ``Quark-antiquark potential in AdS
at one loop," JHEP \textbf{1011} (2010) 079 
[arXiv:1009.3939[hep-th]].
\bibitem{MKA} M. Kruczenski and A. Tirziu, ``Matching the circular
Wilson loop with dual open string solution at 1-loop in
strong coupling," JHEP \textbf{0805} (2008) 064 
[arXiv:0803.0315[hep-th]].
\bibitem{DG} N. Drukker and D.J. Gross, ``An exact prediction
of $\mathcal{N}=4$ SUSYM theory for string theory," 
J. Math. Phys. \textbf{42} (2001) 2896 [hep-th/0010274].
\bibitem{VP} V. Pestun, ``Localization of gauge theory on a 
four-sphere and supersymmetric Wilson loops, " Commun, Math.
Phys. \textbf{313} (2012) 71 [arXiv:0712.2824[hep-th]].
\bibitem{DGR} N. Drukker, S. Giombi, R. Ricci and D. Trancanelli,
``Supersymmetric Wilson loops on $S^3$," JHEP \textbf{0805} 
(2008) 017 [arXiv:0711.3226[hep-th]].
\bibitem{KZ} K. Zarembo, ``Supersymmetric Wilson loops," 
Nucl. Phys. \textbf{B643} (2002) 157 [hep-th/0205160].
\bibitem{GRT} N. Drukker, S. Giombi, R. Ricci and D. Trancanelli,
``Wilson loops: From four-dimensional SYM to two-dimensional YM,"
Phys. Rev. \textbf{D77} (2008) 047901 [arXiv:0707.2699[hep-th]];
V. Branding and N. Drukker, ``BPS Wilson loops in $\mathcal{N}=4$
SYM: Examples on hyperbolic submanifolds of space-time," 
Phys. Rev. \textbf{D79} (2009) 106006 [arXiv:0902.4586[hep-th]];
S. Giombi, V. Pestun and R. Ricci, ``Notes on supersymmetric 
Wilson loops on a two-sphere," JHEP \textbf{1007} (2010) 088 
[arXiv:0905.0665[hep-th]].
\bibitem{VPE} V. Pestun, ``Localization of the four-dimensional
$\mathcal{N}=4$  SYM to a two-sphere and 1/8 BPS Wilson loops,"
JHEP \textbf{1212} (2012) 067 [arXiv:0906.0638[hep-th]].
\bibitem{DF} N. Drukker and and V. Forini, ``Generalized 
quark-antiquark potential at weak and strong 
coupling," JHEP \textbf{1106} (2011) 131 [arXiv:1105.5144[hep-th]].
\bibitem{DGO} N. Drukker, D.J. Gross and H. Ooguri, ``Wilson loops
and minimal surfaces," 
Phys. Rev. \textbf{D60} (1999) 125006 [hep-th/9904191].
\bibitem{MOS} Y. Makkenko, P. Olesen and G.W. Semenoff, 
``Cusped SYM Wilson loop at two loops and beyond," 
Nucl. Phys. \textbf{B748} (2006) 170 [hep-th/0602100].
\bibitem{CH} D. Gorrea, J. Henn, J. Maldacena and A. Sever,
``An exact formula for the radiation of a moving quark in
$\mathcal{N}=4$ super Yang Mills," JHEP \textbf{1206} (2012) 
048 [arXiv:1202.4455[hep-th]].
\bibitem{FGL} B. Fiol, B. Garolera and A. Lewkowycz,
``Exact results for static and radiative fields of a quark
in $\mathcal{N}=4$ super Yang-Mills," JHEP \textbf{1205} (2012) 
093 [arXiv:1202.5292[hep-th]].
\bibitem{CHM} D. Gorrea, J. Henn, J. Maldacena and A. Sever,
``The cusp anomalous dimension at three loops and beyond," 
JHEP \textbf{1205} (2012) 098 [arXiv:1203.1019[hep-th]].
\bibitem{CMS} D. Gorrea, J. Maldacena and A. Sever,
``The quark anti-quark potential and the cusp anomalous dimension
from a TBA equation, "  JHEP \textbf{1208} (2012) 134
[arXiv:1203,1913[hep-th]].
\bibitem{ND} N. Drukker, ``integrable Wilson loops,"
arXiv:1203,1617[hep-th].
\bibitem{GS} N. Gromov and A. Sever,
``Analytic solution of Bremsstrahlung TBA, " JHEP \textbf{1211} 
(2012) 075 [arXiv:1207.5489[hep-th]].
\bibitem{BZ} D. Bykov and K. Zarembo, ``Ladders for Wilson
loops beypnd leading order, " JHEP \textbf{1209} (2012) 057
[arXiv:1206.7117[hep-th]];
J.M. Henn and T. Huber, ``Systematics of the cusp anomalous
dimension, " JHEP \textbf{1211} (2012) 058 
[arXiv:1207.2161[hep-th]];
``The four-loop cusp anomalous dimension and analytic integration
techniques for Wilson line integrals," arXiv:1304.6418[hep-th];
N. Gromov, F. Levkovich-Maslyuk and G. Sizov, ``Analytic
solution of Bremsstrahlung TBA II: Turning on the sphere
angle, " arXiv:1305.1944[hep-th];
M. Beccaria and G. Macorini, ``On a discrete symmetry of the
Bremsstrahlung function in $\mathcal{N} =4$ SYM, "
JHEP \textbf{1307} (2013) 104 [arXiv:1305.4839[hep-th]].
G. Sizov and S. Valatka, ``Algebraic curve for a cusped Wilson
line, " arXiv:1306.2527[hep-th].
\bibitem{FBT}B. Fiol, B. Garolera and G. Torrents, ``Exact momentum 
fluctuations of an accelerated quark in $\mathcal{N} =4$
super Yang-Mills," JHEP \textbf{1306} (2013) 011
[arXiv:1302,6991[hep-th]].
\bibitem{BB} B. Basso, ``An exact slope for AdS/CFT, "
arXiv:1109.3154[hep-th].
\bibitem{KT} M. Kruczenski and A.A. Tseytlin, ``Wilson loops 
T-dual to short strings, " Nucl. Phys. \textbf{B875} (2013)
213 [arXiv:1212.4886[hep-th]].
\bibitem{BT} E.I. Buchbinder and A.A. Tseytlin, ``One-loop
correction to energy of a wavy line string in $AdS_5$, "
arXiv:1309.1581[hep-th].
\bibitem{RKT} R. Kallosh and A.A. Tseytlin,``Simplifying
superstring action on $AdS_5 \times S^5$," JHEP 
\textbf{9810} (1998) 016 [hep-th/9808088].
\bibitem{LAM} L.F. Alday and J.M. Maldacena,``Gluon scattering
amplitudes at strong coupling," JHEP \textbf{0706} (2007) 064
[arXiv:0705.0303[hep-th]].
\bibitem{RTW} R. Ricci, A.A. Tseytlin and M. Wolf,``On
T-duality and integrability for strings on AdS background," 
JHEP \textbf{0712} (2007) 082 [arXiv:0711.0707[hep-th]];
N. Berkovits and J.M. Maldacena,``Fermionic T-duality,
dual superconformal symmetry, and the amplitude/Wilson loop
connection," JHEP \textbf{0809} (2008) 062 
[arXiv:0807.3196[hep-th]];
N. Beisert, R. Ricci, A.A. Tseytlin and M. Wolf,``Dual 
superconformal symmetry from $AdS_5 \times S^5$
superstring integrability," Phys. Rev. \textbf{D78} (2008) 126004
[arXiv:0807.3228[hep-th]].
\bibitem{AM} A. Mikhailov,``Nonlnear waves in AdS/CFT
correspondence, " hep-th/0305196.
\bibitem{HOS} H. Hayashi, K. Okamura, R. Suzuki and B. Vicedo,
``Large winding sector of AdS/CFT," JHEP \textbf{0711} (2007)
033 [arXiv:0709.4033[hep-th]].
\bibitem{FT} S. Frolov and A.A. Tseytlin, ``Semiclassical 
quantization of rotating superstring in $AdS_5 \times S^5$,"
JHEP \textbf{0206} (2002) 007 [hep-th/0204226].


\end{thebibliography}
\end{document}